# CHARACTERIZATION OF ADDITIVE MANUFACTURING MATERIALS FOR STRING ASSEMBLY IN CLEANROOM*

J. Bernardini†, M. Parise, D. Passarelli, FNAL, Batavia, IL, USA

## Abstract

Beamline components, such as superconducting radio frequency cavities and focusing lenses, need to be assembled together in a string while in a cleanroom environment. The present contribution identifies and characterizes materials for additive manufacturing that can be used in a cleanroom. The well known advantages of additive manufacturing processes would highly benefit the design and development of tooling needed for the mechanical support and alignment of string components. Cleanliness, mechanical properties, and leak tightness of the chosen materials are the main focus of this contribution, which also paves the way for the integration of such materials in cryomodule assemblies. Results reported here were obtained in the framework of the PIP-II project at Fermilab.

## INTRODUCTION

The cleanroom assembly [1] of beamline components requires precise connections that are free from particles and leaks [2]. Additionally, each beamline component needs to be individually supported and aligned in relation to the others. This necessitates the design and production of custom tooling that can handle, support, and align these components. For the development of such tooling, materials like stainless steel, aluminum, titanium, and silicone bronze are commonly chosen. After machining and welding, these materials should be electropolished or anodized to ensure smooth and easily cleanable surfaces.

To expedite the development of cleanroom assembly tooling and reduce associated lead time and cost, additive manufacturing techniques such as stereolithography (SLA), fused deposition modeling (FDM), and digital light processing (DLP) could be utilized. These techniques enable the customization of tooling designs according to the unique requirements of each project. However, before utilizing additive manufacturing techniques, it is necessary to qualify the associated materials for cleanroom use. This involves characterizing the mechanical properties and composition of the materials to be considered in the design. To this end, six different materials, including Accura 25 (SLA, 3D systems®, solid density 1.19 g/cm$^3$), Somos WaterShed (SLA, DSM®, solid density 1.12 g/cm$^3$) with and without clear coating, Somos WaterClear (SLA, DSM®, solid density 1.13 g/cm$^3$), Epoxy 82 (DLP, Carbon®, solid density 1.16 g/cm$^3$), and ABS M30i (FDM, Stratasys®), have been procured as equally sized flanges (as shown in Fig. 1) and test samples. After conducting an initial visual inspection, the mechanical properties of Accura 25, Somos WaterShed with clear coating, and ABS M30i have been tested at three different temperatures: 293 K, 77 K (liquid nitrogen), and 4 K (liquid helium). Additionally, the cleanliness, ease of cleaning, and leak-tightness of ACCURA 25 and SOMOS WaterShed with clear coating have been assessed.

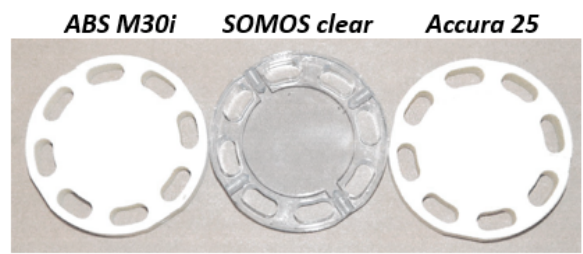

Figure 1: Flanges in ABS M30i, SOMOS watershed with clear coating, Accura 25.

## VISUAL INSPECTION

The flanges underwent a visual inspection to assess their appearance and detect any surface defects or scratches. Flanges made of SOMOS WaterShed and SOMOS WaterClear materials are translucent. Among the samples, the clearest and smoothest surface was achieved through the clear coating applied to SOMOS WaterShed, as shown in Fig. 2. ACCURA 25 flanges consistently exhibit a smooth texture without any visible defects. The unsupported surfaces possess a glossy appearance, while the supported surfaces appear rougher. Epoxy 82 flanges are prone to developing scratches and fractures near corners and sharp edges, as depicted in Fig. 3. Fractures of this nature were observed to emerge as a result of thermal cycling subsequent to the essential photo-polymerization process involved in digital light processing. These defects make Epoxy 82 flanges unsuitable for creating leak-tight connections. ABS M30i flanges display typical rough surfaces resulting from the FDM process. Although not suitable for leak-tight connections, ABS M30i has been subjected to mechanical characterization due to its affordability and accessibility through FDM.

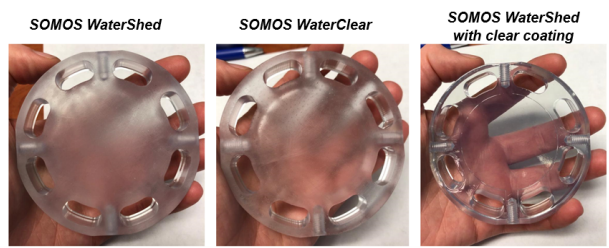

Figure 2: Comparison between clear flanges.

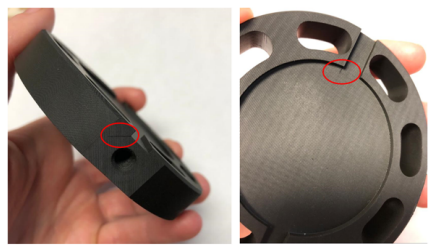

Figure 3: Flanges in Epoxy 82, fractures are visible near corners and sharp edges.

---

* Work supported by Fermi Research Alliance, LLC under Contract No. DEAC02- 07CH11359 with the United States Department of Energy, Office of Science, Office of High Energy Physics.
† jbernard@fnal.gov

Table 1: Tensile Properties

| Material | Temperature [K] | Yield strength [psi] | Tensile Strength [psi] | Elongation [%] |
|---|---|---|---|---|
| SOMOS | 293 | 6540 ± 170 | 7497 ± 201 | 10.00 ± 0 |
|  | 77 | 8270 ± 856 | 8500 ± 838 | 9.33 ± 2.62 |
|  | 4 | 3117 ± 549 | 4027 ± 427 | 2.33 ± 0.47 |
| ACCURA 25 | 293 | 5717 ± 146 | 7067 ± 165 | 22.67 ± 0.47 |
|  | 77 | 4833 ± 1734 | 7877 ± 387 | 7.67 ± 0.47 |
|  | 4 | 8593 ± 25 | 8903 ± 69 | 5.00 ± 0 |
| ABS M30i | 293 | 3180 ± 139 | 3800 ± 29 | 10.33 ± .25 |
|  | 77 | 2450 ± 232 | 2627 ± 239 | 2.33 ± 0.47 |
|  | 4 | 1603 ± 298 | 2167 ± 87 | 2.67 ± 0.47 |

Table 2: Flexural Properties

| Material | Temp [K] | Flexural strength [psi] | Flexural modulus [psi] | Flexural extension [mm] |
|---|---|---|---|---|
| SOMOS | 293 | 11626 ± 99 | 329062 ± 6862 | 0.26 |
|  | 77 | 16855 ± 445 | 825027 ± 45437 | 0.11 |
|  | 4 | 20726 ± 523 | 1629786 ± 13222 | 0.07 |
| ACCURA 25 | 293 | 9643 ± 83 | 240766 ± 5895 | 0.35 |
|  | 77 | 22232 ± 900050 | 11626 ± 23827 | 0.12 |
|  | 4 | 30568 ± 887 | 1533853 ± 30898 | 0.1 |
| ABS M30i | 293 | 8334 ± 566 | 289241 ± 31431 | 0.29 |
|  | 77 | 9445 ± 138 | 854935 ± 202232 | 0.07 |
|  | 4 | 11251 ± 773 | 991164 ± 253235 | 0.07 |

## TEST RESULTS

### Material Composition

The flanges made of ACCURA 25, SOMOS WaterShed with clear coating, and ABS M30i underwent analysis using Pyrolysis/Gas-Chromatography (Py-GC/MS), a method that involves the thermal decomposition of materials in an inert atmosphere. The flanges were subjected to heating in a micro-furnace capable of reaching temperatures up to 1050 °C. During this process, large molecules within the samples undergo cleavage at their weakest points, resulting in the production of smaller and more volatile fragments. These fragments are then separated using gas chromatography. The obtained GC/MS data provides valuable information for identifying individual fragments and determining the structural characteristics of the tested materials.

The GC/MS analysis of the SOMOS WaterShed flange revealed that it primarily consists of vinyl ester resin, which includes long-chain methacrylate and 4,4'-Isopropylidenedicyclohexanol, a non-aromatic derivative of bisphenol-A. Additionally, evidence of a thiol-based photoinitiator was detected in the analysis.

In the case of the ACCURA 25 flange, the GC/MS analysis indicated that it is primarily composed of bisphenol-A epoxy resin. Similar to the SOMOS WaterShed flange, evidence of a thiol-based photoinitiator was also detected.

The GC/MS analysis of the ABS M30i flange confirmed that it consists mainly of acrylonitrile-butadiene-styrene (ABS) material.

### Mechanical Properties

Mechanical properties in terms of yield, tensile, flexural strength, and flexural modulus were measured at ambient (293 K), liquid nitrogen (77 K), and liquid helium (4 K) temperature. The purpose of conducting tests at cryogenic temperatures was to evaluate the suitability of additive manufacturing materials for use in cryomodules.

Results are summarized in Tables 1 and 2. Tests were done on three identical rectangular samples of each material. Samples dimensions are 3.2x12.7x125 mm, tensile tests were performed according to ASTM D638-22 standard, flexural tests were performed according to ASTM D790-

Table 3: Friction Coefficient on Titanium Gr 2

| Material | Side | Static | Kinematic |
|---|---|---|---|
| SOMOS | Side 1 | 0.30 ± 0.02 | 0.29 ± 0.01 |
| ACCURA 25 | Side 1 | 0.42 ± 0.01 | 0.34 ± 0.02 |
|  | Side 2 | 0.19 ± 0.02 | 0.19 ± 0.01 |
| ABS M30i | Side 1 | 0.17 ± 0.01 | 0.15 ± 0.02 |

Table 4: Friction Coefficient on Stainless Steel 304

| Material | Side | Static | Kinematic |
|---|---|---|---|
| SOMOS | Side 1 | 0.32 ± 0.04 | 0.26 ± 0.03 |
| ACCURA 25 | Side 1 | 0.45 ± 0.05 | 0.29 ± 0.02 |
|  | Side 2 | 0.31 ± 0.03 | 0.26 ± 0.01 |
| ABS M30i | Side 1 | 0.28 ± 0.05 | 0.23 ± 0.01 |

17 RL standard. Tensile and flexural tests were done on a 5500R Universal testing machine. For comparison, the minimum yield and tensile strengths of Aluminum 6061-T6 (UNS A96061) at ambient temperature are 42 000 psi and 35 000 psi, respectively. The modulus of elasticity is $10^7$ psi [3] and the density is 2.7 g/cm$^3$.

The measurement of friction coefficient was conducted solely at ambient temperature. This involved sliding rectangular samples against both a Titanium plate and a Stainless Steel 304 plate, following the ASTM D1894-14 standard. In the case of ACCURA 25 samples, the friction coefficient was measured separately for the supported (rough) and unsupported (glossy) sides. However, for the SOMOS WaterShed sample, the clear coating applied eliminated any distinction between the supported and unsupported sides. Notably, no distinction was observed between the upper and lower sides of the ABS M30i sample. Results are summarized in Tables 3 and 4.

*Dielectric Strength*

The dielectric strength was measured by using a GW INSTEK GPT-815 voltage tester, according to S/N EH896585 standard. Results are collected in Table 5. Due to limited input voltage capability (5 kV), minimum value of dielectric strength are provided for ACCURA 25 and SOMOS WaterShed.

Table 5: Dielectric Strength

| Material | Dielectric Strength [kV/cm] |
|---|---|
| SOMOS | >104 |
| ACCURA 25 | >136 |
| ABS M30i | 18.8 |

*Cleaning and Leak Check*

The cleaning procedure performed on the flanges in ACCURA 25 and SOMOS WaterShed includes rinsing in DI water and 99.9% isopropyl alcohol, followed by ultrasonic cleaning, and then dry blowing with filtered nitrogen. A particles counter is used to monitor size and count of particles released by the surface of the flanges as shown in Fig. 4. The particles counter is set to a 7-second cycle with a pumping speed of 75 standard liters per minute (SLPM). The blowing continues until the total number of particulates larger than 0.3 µm is reduced to less than ten in one standard cubic foot of air. This cleaning procedure adheres to the standard protocol implemented at Fermilab for cleaning string components before their assembly in a cleanroom [4]. The time required to achieve a particle count of less than 10, for particles larger than 0.3 µm, in one cubic foot of air varied for the two flanges. Specifically, for the flange made of SOMOS WaterShed with clear coating, this threshold was reached within a range of 1 to 2 minutes. On the other hand, for the flanges composed of ACCURA 25, the process took between 6 and 7 minutes.

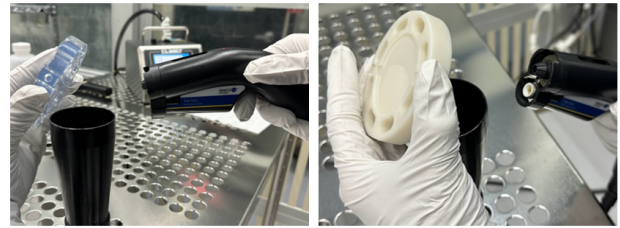

Figure 4: Nitrogen blowing of flanges in Somos Watershed (left) and Accura 25 (right).

The two test flanges were connected to a SS 316 flange by means of M8x1.25 SS 316 rods, Silicone Bronze nuts and, SS 316 washers. The torque used to make each bolted connection was 5 N/m. The SS flange featured a welded nipple for connection with a leak detector. To ensure a secure seal, a Viton o-ring was placed between the flange under testing and the SS 316 flange.

A leak check was conducted with a standard Helium Mass Spectrometer Leak Detector, as shown in Fig. 5. No leak greater than 1 x $10^{-7}$ mbar×liters/sec was detected on both the flanges.

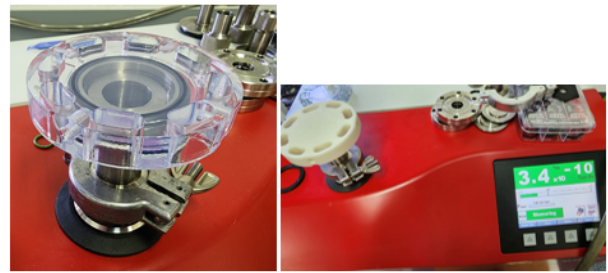

Figure 5: Leak check performed on flanges in SOMOS Watershed and ACCURA 25.

## CONCLUSION

Six different flanges made of materials commonly used in additive manufacturing (Somos WaterShed with and without clear coating, SOMOS WaterClear, ACCURA 25, Epoxy 82, ABS M30i) were obtained and subjected to visual inspection. Based on the results of the visual inspection, ACCURA 25, SOMOS WaterShed, and ABS M30i were identified as the most suitable materials for developing cleanroom string assembly tooling. The mechanical properties of these three materials were characterized at ambient temperature, as well as at liquid nitrogen and liquid helium temperatures. The friction coefficient and dielectric strength were measured solely at ambient temperature.

ACCURA 25 and SOMOS WaterShed flanges were found to be easily cleanable, meeting the cleanliness standards required for string components at Fermilab. These flanges can be utilized to create temporarily leak-tight connections during the assembly of string components.

However, for complete qualification of these materials for use in cryomodules, further characterization is necessary. This includes assessing their thermal properties and radiation hardness, among other relevant factors.

## ACKNOWLEDGEMENTS

The authors would like to thank the personnel of St. Louis Testing Laboratories, Inc. [5] for conducting tests on material characterization and mechanical properties presented in this contribution. Test samples were fabricated at Midwest Prototyping [6], a Protetek company.